\documentclass{article}

\usepackage{spconf}
\usepackage{amssymb,amsthm,amsmath,array}
\usepackage{graphicx}
\usepackage[footnotesize]{caption}
\usepackage[caption=false,font=footnotesize]{subfig}

\usepackage{xspace}
\graphicspath{{Figures/}}
\usepackage[sort&compress, numbers]{natbib}
\usepackage{stmaryrd}
\usepackage{xcolor}

\usepackage{tikz}
\usetikzlibrary{arrows}

\usepackage{pgfplots}
\usepackage{multirow}
\definecolor{mycolor1}{rgb}{0.83,0.83,1}

\newtheorem{prop}{Proposition}
\newtheorem{cor}{Corollary}

\usepackage{color}
\usepackage{hyperref}

\newcommand{\supp}{{\rm supp}\,}

\DeclareMathOperator{\ve}{vec}

\DeclareMathOperator*{\argmin}{arg\,min}

\newcommand{\bs}{\boldsymbol}
\newcommand{\bb}{\mathbb}
\newcommand{\cl}{\mathcal}

\newcommand{\ts}{\textstyle}
\newcommand{\ie}{\emph{i.e.}, }
\newcommand{\eg}{\emph{e.g.}, }

\newcommand{\iid}{%
    \ifmmode
        \mathrm{iid}%
    \else%
        iid\xspace%
    \fi%
}

\newcommand{\whp}{\mbox{w.h.p.\@}\xspace}
\newcommand{\st}{\mbox{s.t.\@}\xspace}

\newcommand{\sq}{\vspace{-2.5mm}}

\newcommand{\myparagraph}[1]
{\ \\[-3mm]
\noindent\textbf{#1}\ }

\begin{document}

\title{\vspace{-15mm}Multilevel Illumination Coding for Fourier Transform Interferometry in Fluorescence Spectroscopy\vspace{-5mm}}
\name{A. Moshtaghpour\textsuperscript{1},
	  L. Jacques\textsuperscript{1} \vspace{-5mm}
	  \thanks{\textsuperscript{1} ISPGroup, ICTEAM, UCLouvain, Belgium. The authors thank P. Antoine and M. Roblin (Lambda-X SA, Nivelles, Belgium) for their help in the acquisition of the FTI measurements. AM is funded by the FRIA/FNRS. LJ is funded by the F.R.S.-FNRS.}}
\address{}
\maketitle

\begin{abstract}
Fourier Transform Interferometry (FTI) is an interferometric procedure for acquiring HyperSpectral (HS) data. Recently, it has been observed that the light source highlighting a (biologic) sample can be coded before the FTI acquisition in a procedure called Coded Illumination-FTI (CI-FTI). This turns HS data reconstruction into a Compressive Sensing (CS) problem regularized by the sparsity of the HS data. CI-FTI combines the high spectral resolution of FTI with the advantages of reduced-light-exposure imaging in biology. 

In this paper, we leverage multilevel sampling scheme recently developed in CS theory to adapt the coding strategy of CI-FTI to the spectral sparsity structure of HS data in Fluorescence Spectroscopy (FS). This structure is actually extracted from the spectral signatures of actual fluorescent dyes used in FS. Accordingly, the optimum illumination coding as well as the theoretical recovery guarantee are derived. We conduct numerous numerical experiments on synthetic and experimental data that show the faithfulness of the proposed theory to experimental observations.
\end{abstract}


\sq
\section{Introduction}                           
\label{sec:intro}           
\sq

Fourier Transform Interferometry (FTI) has received renewed interests in biomedical Fluorescence Spectroscopy (FS) where acquiring HyperSpectral (HS) data with high spectral resolution is crucial to distinguish constituents with slightly different spectral signatures \cite{Leonard2015,Yudovsky2010,lu2014medical}.

\begin{figure}[tb]
	\centering
	\includegraphics[width=0.9\columnwidth]{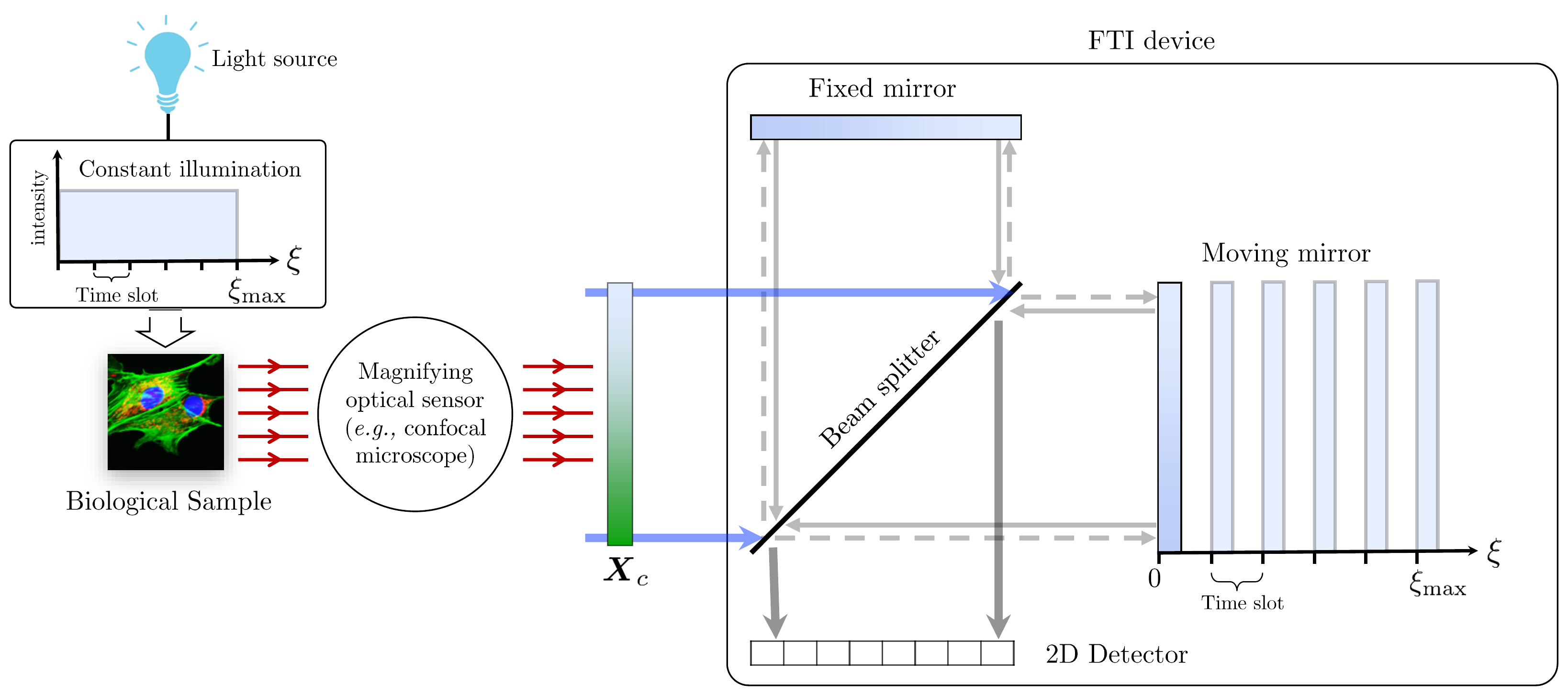}
	\vspace{-2mm}
	\caption{Operating principle of FTI and CI-FTI.}
	\label{fig:fti}
        \vspace{-4mm}
\end{figure}

When it is designed from a Michelson interferometer~\cite{bell2012introductory}, the working principles of FTI for microscopic HS imaging of a biologic sample is explained as follows (see also Fig.~\ref{fig:fti}). A spatially magnified HS light beam, denoted by $\boldsymbol{X}_c$, originating from the highlighted sample (\eg as obtained in confocal microscopy) is first divided by a Beam Splitter (BS). The resulting beam copies are then reflected back to the BS either by a fixed or by a moving mirror, this last element controlling the Optical Path Difference (OPD) of the two beams. After their recombination by the BS, the interferometric intensity of the resulting beam is finally recorded by a 2D imaging sensor. 

Physical optics shows that FTI observations collected in each imager pixel for multiple values of the OPD $\xi \in \bb R$ sample the Fourier transform of the HS volume along the wavenumber domain parameterized by $\nu \in \bb R$, \ie $\xi$ and $\nu$ are (Fourier) dual parameters. As an advantage, the spectral resolution of the HS volume can be increased by enlarging the range of recorded OPD values. However, this increase of resolution is limited by the durability of the fluorescent dyes when exposed to illumination. In fact, over-exposed fluorochromes lose their ability to fluoresce, \ie a phenomenon known as \textit{photo-bleaching} \cite{Ghauharali2000}.

Coded Illumination-FTI (CI-FTI) has been introduced in \cite{moshtaghpour2016,Moshtaghpour2017a} to mitigate the limitation of conventional FTI. By resorting to the theory of Compressive Sensing (CS) \cite{donoho2006compressed,candes2006near} and temporally coding the illumination of the light source, CI-FTI succeeds in reconstructing the target HS volume while minimizing the global light exposure of the observed sample. The authors in \cite{moshtaghpour2018} exploited a type of Variable Density Sampling (VDS) strategy developed in \cite{krahmer2014stable} to recover any HS volume whose spectra are sparsely representable in a wavelet basis.  

This paper proposes to optimize further the illumination coding strategy for FS by leveraging the notion of MultiLevel Sampling (MLS) \cite{Adcock2013} (see Sec. \ref{sec:compr-sens-struct}), a promising VDS extension already applied in Magnetic Resonance Imaging (MRI) and in other FS experiments \cite{roman2014asymptotic}. An optimum multilevel illumination coding scheme is thus established in Sec.~\ref{sec:main-results-cifti} by exploiting the typical structure underlying all spectral sparsity patterns in FS experiments. This entails studying fluorochrome dataset \cite{fluorochromes} and the spectra therein. Conversely to \cite{roman2014asymptotic} where the spectral dimension is scanned sequentially and MLS is applied on the spatial dimension, our approach applies MLS on the Fourier transform of the spectra.

The rest of the paper is structured as follows. We first summarize the recovery guarantee associated to MLS theory in Sec.~\ref{sec:compr-sens-struct}. The new multilevel illumination coding is presented in Sec.~\ref{sec:main-results-cifti}.  Sec.~\ref{sec:numerical-results} demonstrates numerically the power of this approach before concluding the paper.

\textit{Notations:} Vectors and matrices are associated with bold symbols. For a matrix $\bs U = (\bs u_1, \cdots, \bs u_{N_2}) \in \bb C^{N_1 \times N_2}$, $\bs u = \ve(\bs U) := (\bs u_1^\top, \cdots, \bs u_{N_2}^\top)^\top \in \bb C^{N_1N_2}$ denotes the vectorization of $\bs U$. The probability of an event $\cl U$ reads $\bb P(\cl U)$. 3D HS volumes will be either represented by their vector or matrix representations. The $\ell_2$-norm is denoted by $\|\cdot\|$, while other $\ell_p$ norms read $\|\cdot\|_p$ with $1\leq p\leq \infty$. The identity matrix of size $N \times N$ is represented as $\bs I_N$, the index set is $[N]:=\{1, \cdots, N\}$, and $|\cl S|$ is the cardinality of a set $\cl S$. The symbols $\bs \Phi$ and $\bs \Psi$ are reserved for unitary bases, with $\bs\Phi_{\rm DFT}$ and $\bs \Psi_{\rm DHW}$ denoting the 1D discrete Fourier (DFT) and Haar wavelet (DHW) basis , respectively.  For two functions $f$ and $g$, we write $f \lesssim g$ if $f \le c\,g$ for some universal constant $c>0$, and $f \gtrsim g$ if $g \lesssim f$.

\sq
\section{Compressed sensing for signals with structured sparsity}          
\label{sec:compr-sens-struct}
\sq

One branch of CS theory studies the recovery of a signal $\bs x \in \bb C^{N}$ from a vector of measurements $\bs y = \bs P_\Omega\bs \Phi^* \bs x + \bs \eta$~\cite{candes2007sparsity}, where $\Omega = \{\Omega_1,\cdots,\Omega_M\} \subset [N]$ is a set of indices of cardinality $M$, $\bs P_\Omega \in \bb C^{N \times N}$ is a projection operator with $(\bs P_\Omega \bs x)_j = x_j$ if $j \in \Omega$ (and zero otherwise), and $\bs \eta$ is an additive noise on the measurement with bounded power $\|\bs \eta\| \le \varepsilon$. If $\bs x$ is assumed sparse (or well approximated by a sparse representation) in some basis $\bs \Psi$, \ie $\bs x = \bs \Psi \bs s$, with $K:=|\supp (\bs s)| \ll N$, then this signal can be recovered by\vspace{-1mm}
\begin{equation}
  \label{eq:bpdn}
\ts \hat{\bs x} = \argmin_{\bs v \in \bb C^{N}} \|\bs \Psi^*\bs v\|_1 \ \st \ \|\bs y -\bs P_{\Omega} \bs \Phi^*\bs v\| \le \varepsilon, \vspace{-1mm}
\end{equation}
provided $\bs \Phi$ and $\bs \Psi$ respect some incoherent condition~\cite{candes2007sparsity}.  

In this paper, we are concerned in optimizing the sensing procedure above for signals displaying a structured form of sparsity. Following \cite{Adcock2013}, this structure is best captured by splitting the signal sparsity patterns ``in levels'', which in turn will also split similarly the sensing procedure itself. Let us describe this in details.  

For a fixed $r \in \bb N$ we first decompose $[N]$ into $r$ disjoint \textit{sparsity levels} $\cl T :=\{\cl T_1,\cdots,\cl T_r\}$ such that $\bigcup_{\ell=1}^r \cl T_\ell = [N]$. Similarly, $r$ disjoint \textit{sampling levels} are defined as $\cl W := \{\cl W_1,\cdots,\cl W_r\}$ with $\bigcup_{\ell=1}^r \cl W_\ell = [N]$.

Given the parameters $\bs k = (k_1,\cdots,k_r)^\top \in \bb N^r$, a vector $\bs s\in \bb R^N$ is called $(\bs k,\cl T)$-sparse in levels, and we write $\bs s \in \Sigma_{\bs k,\cl T}$, if $|\supp \bs P_{\cl T_\ell}\bs s| \le k_\ell$ for all $\ell \in[r]$. For an arbitrary vector $\bs s$, its $(\bs k,\cl T)$-approximation error is $\sigma_{\bs k,\cl T}(\bs s) := \min\{\|\bs s-\bs z\|_1 : \bs z \in \Sigma_{\bs k,\cl T}\}$. Moreover, given an isometry $\bs U \in\bb C^{N\times N}$, the $t^{\rm th}$ \textit{relative sparsity} is defined by
\vspace{-1mm}
\begin{equation}
\label{eq:relative sparsity}
\ts  K_t(\cl W, \cl T, \bs k) = \max_{\bs z \in \Sigma_{\bs k,\cl T} :\, \|\bs z\|_\infty\le 1} \|\bs P_{\cl W_t} \bs U\bs z\|^2.\vspace{-1mm}
\end{equation}

Given $\bs m = (m_1,\cdots,m_r)^\top \in \bb N^r$, the set $\Omega_{\cl W,\bs m} := \bigcup_{t=1}^{r} \Omega_t$ provides a multilevel sampling scheme, or $(\cl W, \bs m)$-MSS, if, for each $1\leq t\leq r$, $\Omega_t  \subseteq \cl W_t$, $|\Omega_t| = m_t \le |\cl W_t|$, and if the entries of $\Omega_t$ are chosen uniformly at random in $\cl W_t$. Furthermore, the $(t,\ell)^{\rm th}$ \textit{local coherence} of $\bs \Phi$ with respect to $\bs \Psi$ is
\vspace{-2mm}
\begin{equation}
  \label{eq:local coherence}
  \ts \mu_{t,\ell}^{\cl W,\cl T}(\bs \Phi,\bs \Psi) = \sqrt{\mu(\bs P_{\cl W_t}\bs \Phi^*\bs \Psi)\,\mu(\bs P_{\cl W_t}\bs \Phi^*\bs \Psi\bs P_{\cl T_\ell})},\vspace{-2mm}
\end{equation}
where $\mu(\bs U) = \max_{i,j} |U_{i,j}|^2 \in \left[N^{-1},1\right]$ is the local coherence of $\bs U$. Within this context, \eqref{eq:bpdn} is guaranteed to find a good signal estimate in the following sense \cite{Adcock2013}.
\vspace{-2mm}
\begin{prop}[\cite{Adcock2013}]
\label{prop:MLS}
	Let $\Omega = \Omega_{\cl W,\bs m}$ be a $(\cl W,\bs m)$-MSS and $(\bs k,\cl T)$ be any pair such that the following holds: for $0<\epsilon\le {\rm exp}(-1)$, $K = k_1+\cdots+k_r$, and $1\le t \le r$,
	\vspace{-1mm}
	\small
	\begin{equation}
		\label{eq:MLS measurement bound 1}
          \ts	m_t \gtrsim |\cl W_t| \, (\sum_{\ell=1}^{r}\mu_{t,\ell}^{\cl W,\cl T}(\bs \Phi,\bs \Psi) \, k_\ell)\,\log(K\epsilon^{-1})\,\log(N),
		\vspace{-1mm}
	\end{equation}
	\normalsize
	where $m_t \gtrsim \hat{m}_t\,\log(K\epsilon^{-1})\,\log(N)$, and $\hat{m}_t$ is such that 
	\vspace{-1mm}
	\small
	\begin{equation}
          \ts 1\ \gtrsim\ \sum_{t=1}^{r} (\frac{|\cl W_t|}{\hat{m}_t}-1)\ \mu_{t,\ell}^{\cl W,\cl T}(\bs \Phi,\bs \Psi)\, K_t(\cl W,\cl T,\bs k),~~ \ell\in[r].
		\label{eq:MLS measurement bound 2}
		\vspace{-1mm}
	\end{equation}
	\normalsize
	Suppose that $\hat{\bs x}\in \bb C^N$ is a minimizer of \eqref{eq:bpdn}. Then, with probability exceeding $1-\epsilon$, we have
	\vspace{-1mm}
	\small
	\begin{equation}
          \ts	\|\bs x-\hat{\bs x}\| \le \beta_1\,\sigma_{\bs k,\cl T}(\bs \Psi^*\bs x) +\beta_2 \, \varepsilon,\label{eq:MLS error bound}
		\vspace{-1mm}
	\end{equation}
	\normalsize
	for some constant $\beta_1, \beta_2>0$.
\end{prop}
\vspace{-2mm}

\myparagraph{Special sparsity cases:} The acquisition physics of many applications, \eg MRI, tomography, electron microscopy, radio interferometry, and FS using FTI (the purpose of this paper), imposes using the Fourier operator in the sensing model. While~\cite{Adcock2013} has considered the first four applications above in the context of the Haar wavelet sparsity basis~\cite{mallat2008wavelet}, this paper proposes studying (in Prop.~\ref{prop:MLS}) the benefit of two possible sparsity bases for FTI in FS, \ie $\bs \Psi = \bs \Psi_{\rm DHW}$ and $\bs \Psi = \bs \Psi_{\rm DFT}$. For the DHW, we follow the conventions of \cite{roman2014asymptotic} by assigning the sparsity levels to the natural dyadic wavelet levels, also associated with dyadic bands for the sampling levels. For the DFT choice, we impose $\cl W=\cl T$ and symmetric levels (around the DC frequency) with identical cardinality. Table~\ref{tab:definitions} gathers all our level definitions for the two systems.

\vspace{-1mm}
\begin{cor}
\label{cor:F/H and F/F}
In the context of the sparsity and sampling levels defined in Table \ref{tab:definitions}, let us set $\bs \Phi = \bs \Phi_{\rm DFT}$. Each of the two requirements below implies \eqref{eq:MLS measurement bound 1} and \eqref{eq:MLS measurement bound 2} in Prop.~\ref{prop:MLS}:
\small
\vspace{-2mm}
\begin{align}
\label{eq:MLS measurement bound 1, H}
&{\rm \scriptsize (DHW)}\hspace{-3mm}&&\ts m_t\ \gtrsim\ (\sum_{\ell=1}^{r}2^{-|t-\ell|/2} \, k_\ell)\, \log(K\epsilon^{-1})\,\log(N)\\
\label{eq:MLS measurement bound 1, F}
&{\rm \scriptsize (DFT)}\hspace{-3mm}&&\ts m_t\ \gtrsim\ \min\left\{|\cl W_t| , |\cl W_t| \, k_t \, \log(K\epsilon^{-1})\, \log(N)\right\}.
\end{align}
\vspace{-9mm}
\end{cor}
\normalsize
\vspace{-2mm}
\begin{proof}
	See \cite{adcock2016note} for the case of DHW. In the second case, given $\cl W=\cl T$ we can show that $K_t(\cl W,\cl T,\bs k) \le k_t$ and $\mu_{t,\ell}^{\cl W,\cl T}(\bs \Phi_{\rm DFT},\bs \Psi_{\rm DFT})  = \delta_{t,\ell}$, where $\delta_{ij}$ is the Kronecker symbol. Applying these in Prop. \ref{prop:MLS} completes the proof.
	\vspace{-2mm}
\end{proof}
Relation \eqref{eq:MLS measurement bound 1, F} enforces full sampling for the levels where $k_t > 0$. However, in the case that the majority of $k_t$ are zero, the use of DFT sparsity basis will require very few number of measurements (see Sec. \ref{sec:main-results-cifti} and Sec. \ref{sec:numerical-results}).
\vspace{-1mm}
\begin{table}[tb]
\footnotesize
\centering
\scalebox{0.92}{
\begin{tabular}{|l|l|}
\cline{1-2}
&\\[-3mm]
\multirow{4}{*}{\rotatebox[origin=c]{90}{$\bs \Psi_{\rm DHW}$}}&$\cl T_\ell =\{l_{\ell-1}+1,\cdots,l_\ell\}$, $1\le \ell \le r$.\\ 
&$l_0 = 0, l_\ell = 2^\ell$, for $\ell \in [r]$; and $r=\log_2(N)$.\\ \cline{2-2}
&\\[-3mm]
&$\cl W_{t+1}:=\{-n_{t}+1,\cdots, n_{t}\}\backslash \cl W_t$, $1\le t \le r-1$ and $\cl W_1 =\{0,1\}$.\\ 
&$n_0 = 0, n_t = 2^t$, for $t \in [r]$.\\ \cline{1-2}
&\\[-3mm]
\multirow{4}{*}{\rotatebox[origin=c]{90}{$\bs \Psi_{\rm DFT}$}}&$\cl T_\ell =\{-l_{\ell}+1,\cdots, -l_{\ell-1}\}\,\bigcup\,\{l_{\ell-1},\cdots,l_\ell\}$, $1\le \ell \le r$.\\
&$l_0 = 0, l_\ell = \ell \, N \,2^{-q+1}$, for $\ell \in [r]$; and $r = 2^q \ll N$.\\ \cline{2-2}
&\\[-3mm]
&$\cl W_t = \cl T_t$, $1\le t \le r$.\\
&$n_0 = 0, n_t = t\,N\,2^{-q+1}$, for $t \in [r]$.\\ 
\cline{1-2}
\end{tabular}}\vspace{-2mm}
\caption{Definitions of sparsity and sampling levels where $\bs \Phi = \bs \Phi_{\rm DFT}$ and $\bs \Psi$ is indicated in the first column. See Cor. \ref{cor:F/H and F/F}.\vspace{-2mm}}
\label{tab:definitions}
\vspace{-4mm}
\end{table}

\sq
\section{Main Results: CI-FTI in FS}             
\label{sec:main-results-cifti}
\sq

We here focus on the framework of CI-FTI proposed in~\cite{moshtaghpour2016}. We refer the reader to \cite{moshtaghpour2018} for the details on the acquisition principles of FTI and its variations. In short, a vectorized form of the CI-FTI acquisition is modeled as $\bs Y = \bs P_\Omega \bs \Phi_{\rm DFT}^* \bs X + \bs W_{\rm CI}$,
where $\bs W_{\rm CI} := \bs P_\Omega\bs W_{\rm Nyq}$, $\bs X \in \bb R^{N_\xi \times N_p}$ is an HS volume with $N_\xi$ spectral bands and $N_p$ spatial pixels, and $\bs W_{\rm Nyq} =(\bs w_1, \cdots, \bs w_{N_p})$ is a noise matrix corrupting the Nyquist measurements and satisfying $\|\bs w_j\| \le \varepsilon_{\rm Nyq}/\sqrt{N_p}$ for all $j \in [N_p]$. Similarly to \cite{moshtaghpour2018}, in this paper we would like to recover the spectra at all the pixels by solving the following convex optimization problem for all $j \in [N_p]$:
\small
\vspace{-1mm}
\begin{equation}
  \label{eq:cifti sub reconstruction}
  \hspace{-1.5mm}\hat{\bs x}_j = \argmin_{\bs u\in \bb R^{N_\xi}}\ts \|\bs \Psi^*\bs u\|_1 \,\st\,\|\bs D(\bs y_j-\bs P_{\Omega}\bs \Phi_{\rm DFT}^*\bs u_j)\| \le  \alpha\varepsilon_{\rm Nyq},\vspace{-1mm}
\end{equation}
\normalsize
where $\bs D$ and $\alpha$ are defined below according to the selected sensing scheme. As will be clear later, we can show that the error on $\bs x$ is bounded as\vspace{-1mm}
\small
\begin{equation}
  \label{eq:cifti error}
  \ts  \|\bs x-\hat{\bs x}\| \le \beta_1\cdot \sum_{j=1}^{N_p}\sigma_{\bs k, \cl T}(\bs \Psi^*\bs x_j)+\beta_2 \cdot \varepsilon_{\rm Nyq},
  \vspace{-1mm}
\end{equation}
\normalsize
for some $\beta_1, \beta_2>0$ only depending on $|\Omega|, N_\xi$, and $N_p$.

\myparagraph{Initial VDS scheme:} Moshtaghpour et al.~\cite{moshtaghpour2018} propose a VDS scheme for temporal coding of light illumination inspired by~\cite{krahmer2014stable}. They subsample the rows of $\bs \Phi_{\rm DFT}^*$ by selecting $M_\xi \gtrsim K \log^3(K) \log^2(N_\xi)$ OPD indices \iid according to a \emph{pmf} inversely proportional to the OPD magnitude, \ie $\ts p(i) \propto \min\{1, |i-N_\xi/2|^{-1}\}$ for all $i \in [N_\xi]$. The reconstruction procedure is then ensured by 
\eqref{eq:cifti sub reconstruction} by assigning the diagonal matrix $\bs D$ as $D_{ii} = p(\Omega_i)^{-1/2}$, and the reconstruction error is then bounded, with high probability (\whp), as in~\eqref{eq:cifti error}, by setting there $\bs k = (K) \in \bb N$, $\cl T = [N_\xi]$, and the other variables being given in Table \ref{tab:parameters}. 

   \begin{table}[tb]
   \centering
   \footnotesize
   \begin{tabular}{|l|c|c|c|c}
   \cline{1-4}
    Approach                & $\alpha$              & $\beta_1$       & $\beta_2$ &  \\ \cline{1-4}
     &&&&\\[-3mm]
   \cite{moshtaghpour2018} & $\sqrt{M_\xi/N_p}$      & $2/\sqrt{K}$    & $\sqrt{N_p}$     &  \\ \cline{1-4}
     &&&&\\[-3mm]
   This work& $\sqrt{M_\xi/N_\xi N_p}$  & $c$           & $c\cdot\sqrt{M_\xi N_p /N_\xi}$     &  \\ \cline{1-4}
   \end{tabular}\vspace{-2mm}
      \caption{The value of the variables in \eqref{eq:cifti sub reconstruction} and \eqref{eq:cifti error} with respect to different approaches, where $c$ is a constant and $M_\xi=m_1+\cdots + m_r$.\vspace{-2mm}}
      \label{tab:parameters}
      \vspace{-4mm}
   \end{table}
   \normalsize

\myparagraph{FS-driven MLS:} Although this VDS scheme in~\cite{moshtaghpour2018} can be applied on any compressive FTI system (provided the HS data are sparse), boosted reconstruction quality can be reached by leveraging the sparsity structure in levels~\cite{Adcock2013}. Practically, the sparsity pattern of a target signal is unknown. However, a common approach is to consider a class of similar signals and to estimate a sparsity pattern, \ie inclusive for that class of signals. For some examples of this approach see \cite{bastounis2014absence,Adcock2013}. 

\begin{figure}[tb]
   \centering
   \vspace{-2mm}
   \begin{minipage}{\columnwidth}
   \centering
	   \input{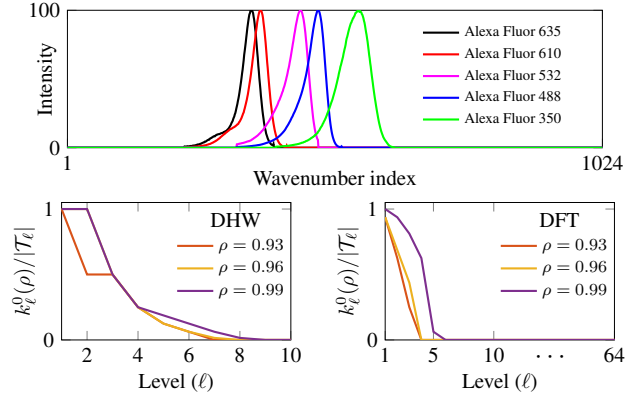}
   \end{minipage}
   \begin{minipage}{0.49\columnwidth}
%
%
\definecolor{mycolor1}{rgb}{0.85000,0.32500,0.09800}%
\definecolor{mycolor2}{rgb}{0.92900,0.69400,0.12500}%
\definecolor{mycolor3}{rgb}{0.49400,0.18400,0.55600}%
\begin{tikzpicture}[scale = 0.8]
\begin{axis}[%
width=1.5in, 
height=0.9in,
at={(0in,0in)},
scale only axis,
xmin=1,
xmax=10,
xlabel={Level ($\ell$)},
xtick={2,4,6,8,10},
xticklabels={2,4,6,8,10},
xlabel style={at = {(0.5,0.04)},font=\fontsize{10}{1}\selectfont},
ticklabel style={font=\fontsize{10}{1}\selectfont},
ymin=0,
ymax=1.05,
ytick={0,1},
yticklabels={0,1},
ylabel={$k_\ell^0(\rho)/|\mathcal{T}_\ell|$},
ylabel style={at = {(0.16,0.5)},font=\fontsize{10}{1}},
axis background/.style={fill=white},
legend style={at={(0.470,0.8)},anchor=north west,legend cell 
align=left,align=left,draw=none,fill=none,font=\fontsize{8}{1}\selectfont}
]
\node[text=black, draw=none] at (rel axis cs:0.76,0.87) {DHW}; 
\addplot [color=mycolor1,solid,line width=1.0pt]
  table[row sep=crcr]{%
1	1\\
2	0.5\\
3	0.5\\
4	0.25\\
5	0.125\\
6	0.0625\\
7	0\\
8	0\\
9	0\\
10	0\\
};
\addlegendentry{$\rho = 0.93$};

\addplot [color=mycolor2,solid,line width=1.0pt]
  table[row sep=crcr]{%
1	1\\
2	1\\
3	0.5\\
4	0.25\\
5	0.125\\
6	0.0625\\
7	0.015625\\
8	0\\
9	0\\
10	0\\
};
\addlegendentry{$\rho = 0.96$};

\addplot [color=mycolor3,solid,line width=1.0pt]
  table[row sep=crcr]{%
1	1\\
2	1\\
3	0.5\\
4	0.25\\
5	0.1875\\
6	0.125\\
7	0.0625\\
8	0.015625\\
9	0\\
10	0\\
};
\addlegendentry{$\rho = 0.99$};

\end{axis}
\end{tikzpicture}%

   \end{minipage}
   \begin{minipage}{0.49\columnwidth}
%
%
\definecolor{mycolor1}{rgb}{0.85000,0.32500,0.09800}%
\definecolor{mycolor2}{rgb}{0.92900,0.69400,0.12500}%
\definecolor{mycolor3}{rgb}{0.49400,0.18400,0.55600}%
\begin{tikzpicture}[scale = 0.8]
\begin{axis}[%
width=1.5in,
height=0.9in,
at={(0in,0in)},
scale only axis,
xmin=1,
xmax=20,
xlabel={Level ($\ell$)},
xtick={1,5,10,20},
xticklabels={1,5,10,64},
xlabel style={at = {(0.5,0.04)},font=\fontsize{10}{1}\selectfont},
ticklabel style={font=\fontsize{10}{1}\selectfont},
ymin=0,
ymax=1.05,
ytick={0,1},
yticklabels={0,1},
ylabel={$k_\ell^0(\rho)/|\mathcal{T}_\ell|$},
ylabel style={at = {(0.16,0.5)},font=\fontsize{10}{1}},
axis background/.style={fill=white},
legend style={at={(0.470,0.8)},anchor=north west,legend cell 
align=left,align=left,draw=none,fill=none,font=\fontsize{8}{1}\selectfont}
]
\node[text=black, draw=none] at (rel axis cs:0.76,0.87) {DFT}; 
\addplot [color=mycolor1,solid,line width=1.0pt]
  table[row sep=crcr]{%
1	0.9375\\
2	0.625\\
3	0.25\\
4	0\\
5	0\\
6	0\\
7	0\\
8	0\\
9	0\\
10	0\\
11	0\\
12	0\\
13	0\\
14	0\\
15	0\\
16	0\\
17	0\\
18	0\\
19	0\\
20	0\\
21	0\\
22	0\\
23	0\\
24	0\\
25	0\\
26	0\\
27	0\\
28	0\\
29	0\\
30	0\\
31	0\\
32	0\\
33	0\\
34	0\\
35	0\\
36	0\\
37	0\\
38	0\\
39	0\\
40	0\\
41	0\\
42	0\\
43	0\\
44	0\\
45	0\\
46	0\\
47	0\\
48	0\\
49	0\\
50	0\\
51	0\\
52	0\\
53	0\\
54	0\\
55	0\\
56	0\\
57	0\\
58	0\\
59	0\\
60	0\\
61	0\\
62	0\\
63	0\\
64	0\\
};
\addlegendentry{$\rho = 0.93$};

\addplot [color=mycolor2,solid,line width=1.0pt]
  table[row sep=crcr]{%
1	0.9375\\
2	0.6875\\
3	0.4375\\
4	0\\
5	0\\
6	0\\
7	0\\
8	0\\
9	0\\
10	0\\
11	0\\
12	0\\
13	0\\
14	0\\
15	0\\
16	0\\
17	0\\
18	0\\
19	0\\
20	0\\
21	0\\
22	0\\
23	0\\
24	0\\
25	0\\
26	0\\
27	0\\
28	0\\
29	0\\
30	0\\
31	0\\
32	0\\
33	0\\
34	0\\
35	0\\
36	0\\
37	0\\
38	0\\
39	0\\
40	0\\
41	0\\
42	0\\
43	0\\
44	0\\
45	0\\
46	0\\
47	0\\
48	0\\
49	0\\
50	0\\
51	0\\
52	0\\
53	0\\
54	0\\
55	0\\
56	0\\
57	0\\
58	0\\
59	0\\
60	0\\
61	0\\
62	0\\
63	0\\
64	0\\
};
\addlegendentry{$\rho = 0.96$};

\addplot [color=mycolor3,solid,line width=1.0pt]
  table[row sep=crcr]{%
1	1\\
2	0.9375\\
3	0.8125\\
4	0.625\\
5	0.0625\\
6	0\\
7	0\\
8	0\\
9	0\\
10	0\\
11	0\\
12	0\\
13	0\\
14	0\\
15	0\\
16	0\\
17	0\\
18	0\\
19	0\\
20	0\\
21	0\\
22	0\\
23	0\\
24	0\\
25	0\\
26	0\\
27	0\\
28	0\\
29	0\\
30	0\\
31	0\\
32	0\\
33	0\\
34	0\\
35	0\\
36	0\\
37	0\\
38	0\\
39	0\\
40	0\\
41	0\\
42	0\\
43	0\\
44	0\\
45	0\\
46	0\\
47	0\\
48	0\\
49	0\\
50	0\\
51	0\\
52	0\\
53	0\\
54	0\\
55	0\\
56	0\\
57	0\\
58	0\\
59	0\\
60	0\\
61	0\\
62	0\\
63	0\\
64	0\\
};
\addlegendentry{$\rho = 0.99$};
\end{axis}

\node[align=left, text=black, draw=none] (a) at (2.78,-0.28) {\fontsize{10}{1}\selectfont $\cdots$}; 
\end{tikzpicture}%
  \end{minipage}\sq
   \caption{The spectra of five fluorochromes (top); The estimated local sparsity ratio for a collection of 38 fluorochrome spectra (bottom).}
   \label{fig:relative sparsity fixed rho}
   \vspace{-4mm}
   \end{figure}

Given a sparsity basis $\bs \Psi$ and the sparsity levels $\cl T$, we estimate the spectral sparsity pattern of the HS volumes in FS as follows. \emph{(i)} We form a dictionary $\bs H \in \bb R^{N_\nu \times N_f}$ by collecting the spectra of $N_f$ fluorochromes commonly used in FS~\cite{Leonard2015}, see Fig.~\ref{fig:relative sparsity fixed rho}-(top). \emph{(ii)} The columns of this dictionary are then represented in $\bs \Psi$ domain, \ie $ \bs 
 H = \bs \Psi \tilde{\bs H}$. \emph{(iii)} For every $\tilde{\bs h}_i$, \ie the $i^{\rm th}$ column of $\tilde{\bs H}$, we define $\pi_{i,j}$ as the index set of the $j$-largest (in absolute value) coefficients of $\tilde{\bs h}_i$. Given $\rho \in [0,1]$, we define the local sparsity of $\tilde{\bs h}_i$ at level $\ell$, as\vspace{-1mm}
\small
$$
\ts k_{i,\ell}(\rho):=|\pi_{i,k(\rho)} \cap \cl T_\ell|,\vspace{-1mm}
$$ 
\normalsize
where $k(\rho) := \min\{n: \|\tilde{\bs h}_{i,\pi_{i,n}}\| \ge \rho \|\tilde{\bs h}_i\|\}$. (iv) Finally, in order to obtain an estimation applicable in FS experiments, we consider the worst local sparsity value among all the fluorochromes included in dictionary $\bs H$, \ie \vspace{-2mm}
\small
$$
\ts k_\ell^0(\rho) := \max \{k_{i,\ell}(\rho): 1 \leq i \leq N_f\} .\vspace{-2mm}
$$\normalsize
For a proper choice of sparsity basis and sparsity levels we should observe that the normalized ratios $k_\ell^0(\rho)/|\cl T_\ell|$ decay rapidly for a fixed $\rho$. 
 
As a proof of concept, we have applied our approach on a collection of $N_f =
38$ spectra of common fluorochromes~\cite{fluorochromes}, frequently used as cell and tissue labels in
FS~\cite{Leonard2015}. This includes the spectra of Alexa Fluors
(Fig.~\ref{fig:relative sparsity fixed rho}-top). We conducted our
test with $\bs
\Psi = \bs\Psi_{\rm DHW}$ and $\bs\Psi = \bs\Psi_{\rm DFT}$, and  
the corresponding sparsity and sensing levels defined in
Table~\ref{tab:definitions} (with $q=6$ for the DFT case). We observe
in Fig.~\ref{fig:relative sparsity fixed rho}-bottom that fluorochrome spectra do display structured sparsity pattern in both DFT and DHW bases. In the DHW basis, local sparsity ratio decreases when the wavelet level increases. Moreover, in the DFT basis, even for a severe $\rho = 0.99$, all the non-zero coefficients are located in the first six levels, \ie in less than 10\% of the coefficients. This compact representation in DFT basis is the main reason for superior HS reconstruction that will be followed in Sec.~\ref{sec:numerical-results}.  

\myparagraph{Validity of this approach:} Our recovery guarantee is valid under the Linear Mixing Model (LMM), which is a common assumption (see \eg \cite{keshava2002spectral}). In LMM, any HS volume $\bs X \in \bb R^{N_\xi \times N_p}$ explained by the spectra matrix $\bs H$ can be modeled as the product $\bs X = \bs H \bs G$, where $\bs G \in \bb R_+^{N_f \times N_p}$ is a mixing matrix with nonnegative entries representing the spatial concentration of each fluorochromes. If $\bs S := \bs \Psi^* \bs X$ is the transformation of (the columns of) $\bs X$ in the spectral basis $\bs \Psi$, then, $\bs S = \tilde{\bs H} \bs G$ and, for each pixel $1\leq j\leq N_p$, the coefficients $\bs s_j$ are mixed as $\bs s_j = \tilde{\bs H} \bs g_j$. Therefore, $\supp \bs s_j \subseteq \bigcup_{i=1}^{N_f} \supp \tilde{\bs h}_i$. Consequently, any sparsity structure shared among the fluorochrome spectra in $\bs H$ is preserved by every spectrum of the HS volume.

\myparagraph{HS Recovery guarantees:} Let us consider a CI-FTI system developed based on one of the two schemes of Table \ref{tab:definitions} and adjusted to the sparsity structure of $\bs H$ as described above. 
Prop.~\ref{prop:MLS} can be invoked to characterize the error bound \eqref{eq:cifti error} on the reconstruction of HS data from \eqref{eq:cifti sub reconstruction} with  $\bs D = \bs I_{N_\xi}$. We select two possible strategies. First, 
we can set $\bs \Psi = \bs \Psi_{\rm DHW}$, and, for each level $t \in [r]$, $m_t$ OPD indices are picked uniformly at random in each sensing level with\vspace{-2mm} 
\small
\begin{equation}
  \label{eq:cifti measurement bound, F/H}
  \ts m_t \gtrsim (\sum_{\ell=1}^{r} 2^{-|t-\ell|/2} \, k_\ell) \, \log(KN_p\epsilon^{-1}) \, \log(N_\xi).\vspace{-2mm}
\end{equation}
\normalsize
Second, we set $\bs \Psi = \bs \Psi_{\rm DFT}$ and suppose that there
exists an integer $r_0 \ll r$ such that $k_{t>r_0} = 0$. 
In this case, \eqref{eq:MLS measurement bound 1, F} imposes to fully
sample the frequencies in the levels $t \in [r_0]$ for any fixed value of $0< \epsilon \le \text{exp}(-1)$. 

Then, from Cor.~\ref{cor:F/H and F/F}, we can easily deduce that, by union bound over all the $N_p$ pixels/spectra, the HS recovery error in \eqref{eq:cifti error} holds with probability exceeding $1-N_p(\epsilon/N_p) = 1- \epsilon$ and with the parameters of Table \ref{tab:parameters}. Note that when the number of measurements increases, $\beta_2$ in \eqref{eq:cifti error} increases too. Thus, taking more measurements does not necessarily imply superior reconstruction. We will see this effect in Sec. \ref{sec:numerical-results}.

\sq
\section{Numerical results}                    
\label{sec:numerical-results}
\sq

Let us now evaluate the performance of CI-FTI using the two FS-driven approaches. All HS data were reconstructed by solving \eqref{eq:cifti sub reconstruction} with the SPGL1 toolbox~\cite{spgl1:2007}. 

The first experiment traces the phase transition curves of successful recovery of a single spectrum with respect to the measurement ratio $M_\xi/N_\xi$ in the noiseless case. The measurements are formed as $\bs y = \bs P_\Omega \bs \Phi_{\rm DFT}^* \bs x$, where $\bs x := \bs \Psi \bs H \bs g$ is a synthetic spectrum resulting from the linear mixing of $N_f$ spectra, as realized by $\bs g \in [0,1]^{N_f}$ with $g_i \sim_{\iid} \cl U([0,1])$. In our new approaches the set $\Omega$ is generated depending on the local sparsity values (see Sec.~\ref{sec:main-results-cifti}). Fig.~\ref{fig:sim1} shows the probability of successful reconstruction over 100 independent realizations of $\Omega$ and $\bs g$. The improvement of the first approach is due to the structured sparsity exploited for the optimum subsampling strategy. In addition, since the spectra are highly compressible in DFT basis, the successful recovery rate is boosted for the second approach.
  \begin{figure}[tb]
  \centering
%
%
\definecolor{mycolor1}{rgb}{0.85000,0.32500,0.09800}%
\begin{tikzpicture}[scale = 0.9]
\begin{axis}[%
width=3in,
height=1.5in,
at={(0in,0in)},
scale only axis,
xmin=0,
xmax=1,
xlabel={Measurement ratio $(M_\xi/N_\xi)$},
xtick={0.1,0.2,0.3,0.4,0.5,0.6,0.7,0.8,0.9,1},
xticklabels={0.1,0.2,0.3,0.4,0.5,0.6,0.7,0.8,0.9,1},
xlabel style={at = {(0.5,0.02)},font=\fontsize{10}{1}\selectfont},
ticklabel style={font=\fontsize{10}{1}\selectfont},
ymin=0,
ymax=103,
ytick={0,20,40,60,80,100},
yticklabels={0,0.2,0.4,0.6,0.8,1},
ylabel={\small $\mathbb{P}(\|\bs x-\hat{\bs x}\|^2/\|\bs x\|^2 \le 10^{-4})$},
ylabel style={at = {(0.05,0.5)},font=\fontsize{10}{1}},
axis background/.style={fill=white},
legend style={at={(0.5,0.4)},anchor=north west,legend cell 
align=left,align=left,draw=none,fill=none,font=\fontsize{8}{1}\selectfont}
]
\addplot [color=blue,solid,line width=1.0pt,mark=diamond,mark options={solid}]
  table[row sep=crcr]{%
0.05	0\\
0.1	0\\
0.2	0\\
0.3	0\\
0.4	98\\
0.5	100\\
0.6	100\\
0.7	100\\
0.8	100\\
0.9	100\\
1	100\\
};
\addlegendentry{New approach 1: DHW};

\addplot [color=black,solid,line width=1.0pt,mark=o,mark options={solid}]
  table[row sep=crcr]{%
0.05	1\\
0.1	100\\
0.2	100\\
0.3	100\\
0.4	100\\
0.5	100\\
0.6	100\\
0.7	100\\
0.8	100\\
0.9	100\\
1	100\\
};
\addlegendentry{New approach 2: DFT};

\addplot [color=mycolor1,solid,line width=1.0pt,mark=+,mark options={solid}]
  table[row sep=crcr]{%
0.05	0\\
0.1	0\\
0.2	0\\
0.3	0\\
0.4	29\\
0.5	98\\
0.6	100\\
0.7	100\\
0.8	100\\
0.9	100\\
1	100\\
};
\addlegendentry{Initial approach};

\end{axis}
\end{tikzpicture}%
    \vspace{-4mm}
  \caption{Comparison of the successful recovery rate with the proposed approaches and the initial approach in \cite{moshtaghpour2018}. }
  \label{fig:sim1}
  \vspace{-4mm}
  \end{figure}
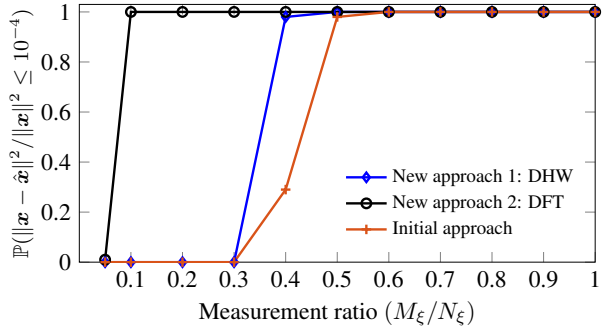
 \begin{figure}[tb]
\begin{minipage}{1.01\linewidth}
  \begin{minipage}{0.31\columnwidth}
  \centering
%
%
\begin{tikzpicture}

\begin{axis}[%
width=1in,
height=1in,
at={(0in,0in)},
scale only axis,
axis on top,
xmin=0.5,
xmax=128.5,
y dir=reverse,
ymin=0.5,
ymax=128.5,
hide axis
]
\addplot [forget plot] graphics [xmin=0.5,xmax=128.5,ymin=0.5,ymax=128.5] {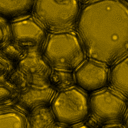};
\node[fill = none,text=white, draw=none] at (rel axis cs:0.26,0.94) {\fontsize{6}{1}\selectfont \textbf{(a) Reference}}; 
\filldraw[color = white, fill = none, thick] (94.5,32.5)rectangle (126.5,64.5);
\end{axis}
\end{tikzpicture}%
  \end{minipage}
  \begin{minipage}{0.31\columnwidth}
  \centering
%
%
\begin{tikzpicture}

\begin{axis}[%
width=1in,
height=1in,
at={(0.729in,0in)},
scale only axis,
axis on top,
xmin=96.5,
xmax=128.5,
y dir=reverse,
ymin=32.5,
ymax=64.5,
hide axis
]
\addplot [forget plot] graphics [xmin=0.5,xmax=128.5,ymin=0.5,ymax=128.5] {Fig_Spatial_ref_DFT-1.png};
\node[fill = none,text=white, draw=none] at (rel axis cs:0.13,0.94) {\fontsize{6}{1}\selectfont \textbf{4x (a)}}; 

\end{axis}
\end{tikzpicture}%
  \end{minipage}
 \begin{minipage}{0.31\columnwidth}
 		  \hspace{-2mm}
		  \begin{minipage}{0.31\columnwidth}
		  \centering
		  \input{Figures/Fig_Sim2_Spectrum_ref.tex}
		  \end{minipage}
		  \\ \\
		  \begin{minipage}{0.31\columnwidth}
		  \centering
		  \end{minipage}
  \end{minipage}
\end{minipage}
\begin{minipage}{1.01\linewidth}
  \begin{minipage}{0.31\columnwidth}
  \centering
%
%
\begin{tikzpicture}

\begin{axis}[%
width=1in,
height=1in,
at={(0in,0in)},
scale only axis,
axis on top,
xmin=0.5,
xmax=128.5,
y dir=reverse,
ymin=0.5,
ymax=128.5,
hide axis
]
\addplot [forget plot] graphics [xmin=0.5,xmax=128.5,ymin=0.5,ymax=128.5] {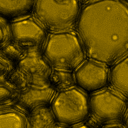};

\node[fill = none,text=white, draw=none] at (rel axis cs:0.37,0.94) {\fontsize{6}{1}\selectfont \textbf{(b) New approach 1}}; 
\filldraw[color = white, fill = none, thick] (94.5,32.5)rectangle (126.5,64.5);
\end{axis}
\end{tikzpicture}%
  \end{minipage}
  \begin{minipage}{0.31\columnwidth}
  \centering
%
%
\begin{tikzpicture}

\begin{axis}[%
width=1in,
height=1in,
at={(0.729in,0in)},
scale only axis,
axis on top,
xmin=96.5,
xmax=128.5,
y dir=reverse,
ymin=32.5,
ymax=64.5,
hide axis
]
\addplot [forget plot] graphics [xmin=0.5,xmax=128.5,ymin=0.5,ymax=128.5] {Fig_Spatial_DHW-1.png};

\node[fill = none,text=white, draw=none] at (rel axis cs:0.13,0.94) {\fontsize{6}{1}\selectfont \textbf{4x (b)}}; 
\end{axis}
\end{tikzpicture}%
  \end{minipage}
  \begin{minipage}{0.31\columnwidth}
   		  \hspace{-2mm}
		  \begin{minipage}{0.31\columnwidth}
		  \centering
		  \input{Figures/Fig_Sim2_Spectrum_DHW.tex}
		  \end{minipage}
		  \\ \\
		  \begin{minipage}{0.31\columnwidth}
		  \centering
%
%
\begin{tikzpicture}
\begin{axis}[%
width=1in,
height=0.2in,
at={(0in,0in)},
scale only axis,
axis on top,
xmin=0.5,
xmax=1024.5,
y dir=reverse,
ymin=0.5,
ymax=1.5,
ytick={\empty},
hide axis
]
\addplot [forget plot] graphics [xmin=0.5,xmax=1024.5,ymin=0.5,ymax=1.5] {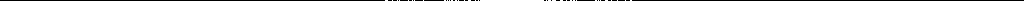};
\end{axis}
\end{tikzpicture}%
		  \end{minipage}
  \end{minipage}
\end{minipage}
\begin{minipage}{1.01\linewidth}
  \begin{minipage}{0.31\columnwidth}
  \centering
%
%
\begin{tikzpicture}

\begin{axis}[%
width=1in,
height=1in,
at={(0.729in,0in)},
scale only axis,
axis on top,
xmin=0.5,
xmax=128.5,
y dir=reverse,
ymin=0.5,
ymax=128.5,
hide axis
]
\addplot [forget plot] graphics [xmin=0.5,xmax=128.5,ymin=0.5,ymax=128.5] {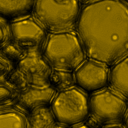};

\node[fill = none,text=white, draw=none] at (rel axis cs:0.37,0.94) {\fontsize{6}{1}\selectfont \textbf{(c) New approach 2}}; 
\filldraw[color = white, fill = none, thick] (94.5,32.5)rectangle (126.5,64.5);
\end{axis}
\end{tikzpicture}%
  \end{minipage}
  \begin{minipage}{0.31\columnwidth}
  \centering
%
%
\begin{tikzpicture}

\begin{axis}[%
width=1in,
height=1in,
at={(0.729in,0in)},
scale only axis,
axis on top,
xmin=96.5,
xmax=128.5,
y dir=reverse,
ymin=32.5,
ymax=64.5,
hide axis
]
\addplot [forget plot] graphics [xmin=0.5,xmax=128.5,ymin=0.5,ymax=128.5] {Fig_Spatial_DFT-1.png};

\node[fill = none,text=white, draw=none] at (rel axis cs:0.13,0.94) {\fontsize{6}{1}\selectfont \textbf{4x (c)}}; 
\end{axis}
\end{tikzpicture}%
  \end{minipage}
 \begin{minipage}{0.31\columnwidth}
  		  \hspace{-2mm}
		  \begin{minipage}{0.31\columnwidth}
		  \centering
		  \input{Figures/Fig_Sim2_Spectrum_DFT.tex}
		  \end{minipage}
		  \\ \\
		  \begin{minipage}{0.31\columnwidth}
		  \centering
%
%
\begin{tikzpicture}
\begin{axis}[%
width=1in,
height=0.2in,
at={(0in,0in)},
scale only axis,
axis on top,
xmin=0.5,
xmax=1024.5,
y dir=reverse,
ymin=0.5,
ymax=1.5,
ytick={\empty},
hide axis
]
\addplot [forget plot] graphics [xmin=0.5,xmax=1024.5,ymin=0.5,ymax=1.5] {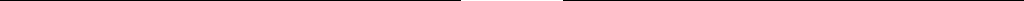};
\end{axis}
\end{tikzpicture}%
		  \end{minipage}
  \end{minipage}
\end{minipage}
\begin{minipage}{1.01\linewidth}
  \begin{minipage}{0.31\columnwidth}
  \centering
%
%
\begin{tikzpicture}

\begin{axis}[%
width=1in,
height=1in,
at={(0.729in,0in)},
scale only axis,
axis on top,
xmin=0.5,
xmax=128.5,
y dir=reverse,
ymin=0.5,
ymax=128.5,
hide axis
]
\addplot [forget plot] graphics [xmin=0.5,xmax=128.5,ymin=0.5,ymax=128.5] {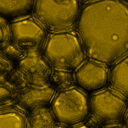};

\node[fill = none,text=white, draw=none] at (rel axis cs:0.37,0.94) {\fontsize{6}{1}\selectfont \textbf{(d) Initial approach}}; 
\filldraw[color = white, fill = none, thick] (94.5,32.5)rectangle (126.5,64.5);
\end{axis}
\end{tikzpicture}%
  \end{minipage}
  \begin{minipage}{0.31\columnwidth}
  \centering
%
%
\begin{tikzpicture}

\begin{axis}[%
width=1in,
height=1in,
at={(0.729in,0in)},
scale only axis,
axis on top,
xmin=96.5,
xmax=128.5,
y dir=reverse,
ymin=32.5,
ymax=64.5,
hide axis
]
\addplot [forget plot] graphics [xmin=0.5,xmax=128.5,ymin=0.5,ymax=128.5] {Fig_Spatial_VDS-1.png};

\node[fill = none,text=white, draw=none] at (rel axis cs:0.13,0.94) {\fontsize{6}{1}\selectfont \textbf{4x (d)}}; 
\end{axis}
\end{tikzpicture}%
  \end{minipage}
 \begin{minipage}{0.31\columnwidth}
  		  \hspace{-2mm}
		  \begin{minipage}{0.31\columnwidth}
		  \centering
		  \input{Figures/Fig_Sim2_Spectrum_VDS.tex}
		  \end{minipage}
		  \\ \\
		  \begin{minipage}{0.31\columnwidth}
		  \centering
%
%
\begin{tikzpicture}
\begin{axis}[%
width=1in,
height=0.2in,
at={(0in,0in)},
scale only axis,
axis on top,
xmin=0.5,
xmax=1024.5,
y dir=reverse,
ymin=0.5,
ymax=1.5,
ytick={\empty},
hide axis
]
\addplot [forget plot] graphics [xmin=0.5,xmax=1024.5,ymin=0.5,ymax=1.5] {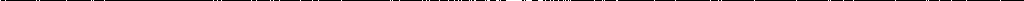};
\end{axis}
\end{tikzpicture}%
		  \end{minipage}
  \end{minipage}
\end{minipage}\sq
\caption{The reconstructed HS volumes. (left) The spatial maps at 594 nm. (right) The spectra at the center pixel, here restricted to the first 200 indices of wavenumber axis. The coding pattern is shown in the last column. New approach 2 results in smoother and less noisy reconstruction.}
\label{fig:sim2_spatial}
\vspace{-4mm}
\end{figure}

The second experiment consists in simulating CI-FTI measurements from actual FTI measurements recorded at Nyquist regime. The full description of this experiment is explained in \cite{moshtaghpour2018}. In short, we observed a thin layer of a cell, \ie Convallaria, lily of the valley, cross section of rhizome with concentric vascular bundles. The Nyquist-FTI measurements of size $(N_\xi,N_x,N_y)=(1024,128,128)$ was obtained at current level 700 mA. We formed CI-FTI measurement by subsampling 10\% of the Nyquist-FTI measurements using the projection $\bs P_\Omega$. The reference HS volume, in this test, is the one reconstructed from the Nyquist sensing.

The reconstructed HS volumes are illustrated in Fig.~\ref{fig:sim2_spatial}. Recall that in CI-FTI the spatial dimension of the measurements is not subsampled; hence, the spatial configuration of the specimen is preserved. We have thus to assess the existence of the noise elements in the reconstructed spatial maps as well as the spectra. The former is clear in the second column. The reconstructed spectra witness the superiority of the two proposed approaches. As mentioned before, increasing the number of measurements consequences in more noise terms in the reconstructed spectra. This can be seen in the reference spectra. However, omitting the noise terms, we can see the shape of the reference spectrum. This shape is accurately preserved by our second approach, thanks to the proper choice of sparsity basis. Furthermore, as a result of subsampling few coefficients, the noise terms are significantly reduced.
 

\sq
\section{Conclusion}                             
\label{sec:conclusion}            
\sq

We have presented new coding designs for CI-FTI that are adapted to FS. These schemes are derived from the notions of sparsity in levels, multilevel sampling, and local coherence in levels \cite{Adcock2013}. For our application, the common sparsity pattern among different fluorochrome spectra has been extracted. Our conclusion is that adapting CI-FTI with the corresponding multilevel illumination coding improves the quality of the reconstructed spectra. Extending this work to coded aperture-FTI, \ie spatio-temporal FTI coding ~\cite{Moshtaghpour2017a}, will be the scope of future research. 

\bibliographystyle{IEEEtran}        
\bibliography{IEEEabrv,refs}

\end{document}